\documentclass[preprint,preprintnumbers,amsmath,amssymb,prb]{revtex4}
\usepackage{amsmath}
\usepackage{graphicx}
\usepackage{dcolumn}
\usepackage{bm}
\usepackage{graphics}
\usepackage{epsfig}
\usepackage{subfigure}
\usepackage{color}

\newcommand\eq{\begin{equation}}
\newcommand\be{\begin{equation}}
\newcommand\eeq{\end{equation}}
\newcommand\ee{\end{equation}}
\newcommand\ar{\begin{eqnarray}}
\newcommand\ear{\end{eqnarray}}

\newcommand{\ii}{{\rm i}}

\def\bea{\begin{eqnarray}}
\def\eea{\end{eqnarray}}

\begin{document}
\newcommand{\nm}{\mbox{ nm}}
\newcommand{\micron}{\mbox{ $\mu$m}}
\newcommand{\fm}{\mbox{ fm}}
\newcommand{\THz}{\mbox{ THz}}
\newcommand{\watt}{\mbox{ W}}
\newcommand{\m}{\mbox{ m}}
\newcommand{\mm}{\mbox{ mm}}
\newcommand{\km}{\mbox{ km}}
\newcommand{\cm}{\mbox{ cm}}
\newcommand{\ps}{\mbox{ ps}}
\newcommand{\seconds}{\mbox{ s}}
\newcommand{\minutes}{\mbox{ min}}
\newcommand{\volts}{\mbox{ V}}
\newcommand{\MeV}{\mbox{ MeV}}
\newcommand{\eV}{\mbox{ eV}}
\newcommand{\meV}{\mbox{ meV}}
\newcommand{\rad}{\mbox{ radians}}

\title{Analysis of surface plasmon excitation at terahertz frequencies with highly-doped graphene sheets via attenuated total reflection}

\author{Choon How Gan}

\affiliation{Electronics and Photonics Department,\\
A*STAR Institute of High Performance Computing, 138632, Singapore}

\begin{abstract}
Excitation of surface plasmons supported by doped graphene sheets at terahertz frequencies is investigated numerically. To alleviate the momentum mismatch between the highly-confined plasmon modes and the incident radiation, it is proposed to increase the surface conductivity of graphene through high doping levels or with few-layer graphene. 
For currently achievable doping levels, our analysis shows that surface plasmons on monolayer graphene may be excited at operating frequencies up to about $10 \THz \,(\sim 41.3 \meV)$ with a high-index coupling prism, and higher frequencies/energies are possible for few-layer graphene. These highly-confined surface modes are promising for sensing and waveguiding applications in the terahertz regime.
\end{abstract}

\maketitle

Electromagnetic fields associated with surface plasmons (SPs) guided by highly-conductive metal surfaces at midinfrared (MIR) and terahertz (THz) frequencies are extremely unconfined; they are plane-wave like and spread ubiquitously in the interfacing dielectric medium. This poorly-confined nature of SPs propagating along near-perfect conductors may be understood by inspection of its wavenumber $k_{sp}$. At a metal-dielectric interface~\cite{raether}, $k_{sp} = k_0 \sqrt{\epsilon_d\epsilon_m/(\epsilon_d + \epsilon_m)}$, where $k_0$ is the wavenumber in free-space, and $\epsilon_d$ and $\epsilon_m$ are the dielectric constants of the dielectric and metal, respectively. 
For the simple half-space problem, the transverse wavenumber of the SP in the dielectric is given by $k_{zd}^2 = k_0^2 \epsilon_d - k_{sp}^2 = k_0^2 \epsilon_d^2/(\epsilon_d + \epsilon_m)$. 
At visible and near-infrared frequencies, $|\Re \epsilon_m| >> \Im \epsilon_m$ (typical of the noble metals), the imaginary part of $k_{zd}$ dominates and the field is well-confined. 
At THz frequencies however, $|\Re \epsilon_m|$ and $\Im \epsilon_m$ are both large with comparable magnitudes. Typically $|\epsilon_m| >> \epsilon_d$, $k_{sp} \sim k_0 \sqrt{\epsilon_d}$, and $k_{zd}^2 \sim k_0^2 \epsilon_d^2/\epsilon_m$, the imaginary part of $k_{zd}$ is relatively small leading to a long decay length in the dielectric. 

To circumvent the poor confinement at long wavelengths, tunable SP-like modes supported on metal surfaces corrugated with nanoholes or nanorods, and on multi-layered metamaterial systems have been proposed~\cite{spoof,kabashin,ganol}.     
More recently, studies have shown that graphene, a single layer of carbon atoms gathered in honeycomb lattice, can also support 
well-confined SP modes at MIR and THz frequencies~\cite{engheta,koppens,ganprb85,bwangapl100}. As its electronic transport properties can be readily tuned by the application of a gate voltage~\cite{novo666,thongapl100}, graphene structures present an attractive alternative as platforms for supporting SPs.  
However, due to the relatively large momentum mismatch between the SPs and the light photons, their excitation on graphene remains a challenging issue. 
Proposed approaches to overcome the mismatch include the patterning of graphene into an array of nanoribbons~\cite{junatnano}, and probing with a nano-sized metalized tip in near-field nanoscopy~\cite{basovnano,jchenarxiv}. In this letter, 
we investigate the excitation of SPs on highly-doped graphene sheets with attenuated total reflection (ATR) via the Otto geometry~\cite{raether,sambles}. 
As shown in Fig.~1, graphene is deposited on a substrate of refractive index $n_1$, which is separated from the coupling prism ($n_p$) by a small gap $d$ ($n_2$). In the following, the substrate is taken to be nonpolar so that effects of remote phonon scattering may be neglected~\cite{fratiniprb77}.
Unlike in metallic thin films where the SP dispersion splits into two branches, monolayer graphene (MLG) sandwiched between two dielectric media supports a 
single bound SP mode~\cite{ganprb85}. For well-confined SP modes supported on MLG, the associated magnetic field ($H$) and normal component of the 
electric field ($E_z$) change sign across the graphene sheet on account of the continuity of the tangential electric field in the two neighboring media with positive permittivity values. In addition to highly-doped MLG, few-layer graphene (FLG) will be considered in our attempt to reduce the momentum mismatch. By taking individual graphene sheet as a non-interacting monolayer, the optical conductivity of the FLG~\cite{ferrarinl,xianatnano} is $N\sigma$, where $N$ is the number of layers ($N < 6$) and $\sigma$ is the conductivity of the MLG. As will be shown shortly, increasing the conductivity effectively reduces the SP wavenumber. 

\begin{figure}[tbp]
\centering
\epsfig{file=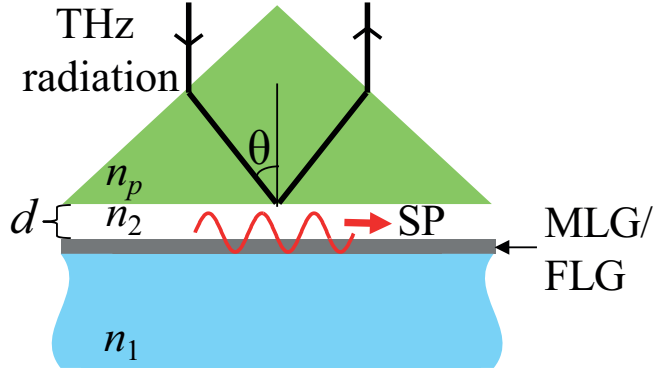, width=3.375 in}
\caption{Exciting SPs on MLG/FLG with the Otto geometry. }
\label{fig:f1}
\end{figure}

Let us model each graphene monolayer as a surface conducting sheet. Near to the THz regime, intraband scattering dominates in highly-doped graphene and its 
conductivity takes on a Drude-like~\cite{engheta,koppens,ganprb85,ranaapl93} form $\sigma \approx \ii e^2 \mu/[\pi \hbar^2 (\omega + \ii \tau^{-1})]$. 
Here $\mu \approx \hbar v_F \sqrt{N_c \pi}$ is the  Fermi-level, 
$v_F \approx 10^6~{\rm m/s}$ is the Fermi velocity, $N_c$ is the carrier density,
$\omega = k_0c$ is the angular frequency with $c$ the speed of light in vacuum,
$\tau$ is a phenomenological electron relaxation time,
and the local limit $\omega >> \tau^{-1}, k_{sp}v_f$ is assumed. 
By matching the boundary conditions for the $n_1$-MLG-$n_2$ system (Fig.~1), the SP dispersion may be derived as~\cite{ganprb85,hansonjap103}
\eq\label{spdispigi}
\frac{n_1^2}{k_{z1}} + \frac{n_2^2}{k_{z2}} + \frac{\sigma}{\omega\epsilon_0} = 0 \,.
\eeq
If $\Re k_{sp} >> \Im k_{sp}$ and $|k_{sp}|^2 >> k_i^2$, where $k_i = n_ik_0$ and $i = \{1, 2\}$, the transverse wavenumber  of the SP is $k_{zi} = \ii \sqrt{k_{sp}^2 - k_i^2 } \approx \ii k_{sp}$. Substituting $k_{zi} = \ii k_{sp}$ into the above 
dispersion relation yields the approximation~\cite{ganprb85,jablanprb80} 
\eq\label{spdispapp}
k_{sp} \approx \frac{\pi\epsilon_0\hbar^2}{e^2} \frac{(n_1^2 + n_2^2) (\omega^2 + \ii\omega\tau^{-1})}{\mu} \,. 
\eeq
From this expression, it is readily seen that the SP wavenumber decreases with increasing $\mu$ or $\sigma$. 
Figure~2 shows the SP dispersion for the MLG  as a function of frequency $f$ in the THz regime and as a function of the Fermi-Level $\mu$. 
For all following calculations, $\tau$ is taken to be $1 \ps$ to account 
for scattering loss from acoustic phonons~\cite{hansonjap103,jablanprb80}, and the refractive indices $n_1$ and $n_2$ are taken to be 1.6 and 1, respectively.  
In Fig.~2(a), the Fermi-level is taken to be $\mu = 0.9 \eV$, which corresponds to $N_c \approx 6 \times 10^{13} \cm^{-2}$, within the doping 
levels achievable in experiments~\cite{efetov,fwangnat471}. 
The decrease in the effective mode index of the SP ($n_{sp} = k_{sp}/k_0$) as 
the frequency decreases is evident. 
The data also shows that SPs can propagate for hundreds of 
microns ($L = \lambda/(4\pi\Im n_{sp})$) in the THz regime with effective wavelengths 2-6 times smaller than the free-space wavelength $\lambda = 2\pi/k_0$. 
For $f  = 5 \THz$, the dependence of the SP dispersion on $\mu$ is shown in Fig.~2(b), from which the decrease of $n_{sp}$ with increased $\mu$ is clearly observed. 
The relations $\Re n_{sp} \propto \omega, \mu^{-1}$ and $L_{sp} \propto \omega^{-1}, \mu$ derived from the 
approximate form Eq.~\eqref{spdispapp} (dashed curves in Fig.~2) is in good agreement with the general trends of the SP dispersion.  
However, as seen from Fig.~2, the approximated dispersion relation~\eqref{spdispapp} starts to deviate significantly from the exact form (Eq.~\eqref{spdispigi})
for frequencies $f \lesssim 3$ THz 
or Fermi-levels $\mu \gtrsim 1.6 \eV$, especially for the imaginary part of the dispersion. 
Beyond these limits, the imaginary part of $k_{zi}$ is no longer the dominant term, leading to the inaccuracies in the approximation~\eqref{spdispapp}.

\begin{figure}[]
\centering
\epsfig{file=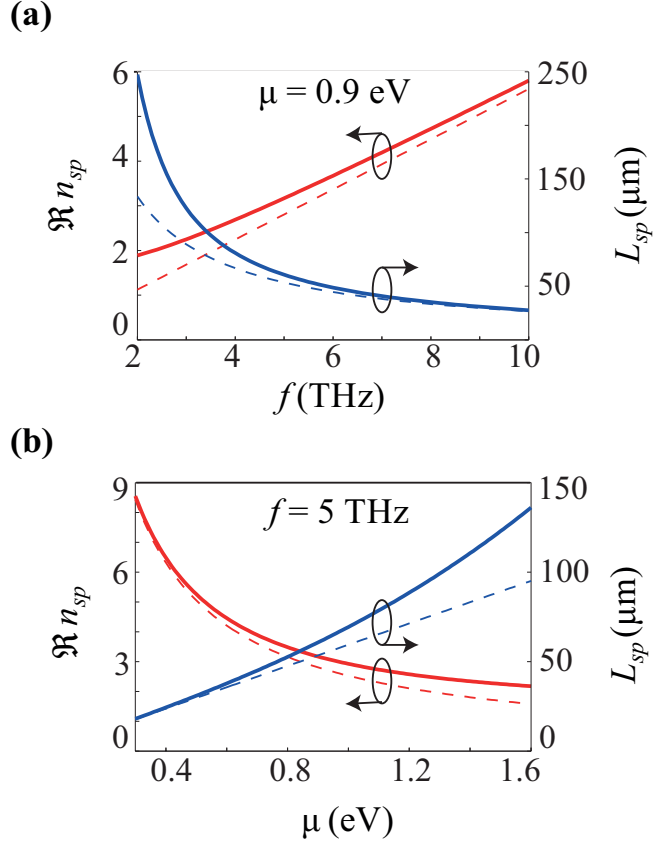, width=3.375 in}
\caption{SP dispersion on MLG as a function of (a) frequency,  and (b) Fermi-level. 
Solid curves are results from Eq.~\eqref{spdispigi}, and dashed curves are calculated from the approximation in Eq.~\eqref{spdispapp}. }
\label{fig:f2}
\end{figure}

The excitation of SPs with the MLG or FLG in the Otto geometry is demonstrated with the calculated angular reflection 
spectra in Fig.~3, taking the operating frequency to be $f = 5 \THz \,(\hbar\omega = 20.7 \meV)$.
In our calculations performed with the transfer-matrix method~\cite{yariv}, 
it is taken that each graphene sheet has effective thickness~\cite{engheta,koppens} $t_g \sim 0.5 \nm$ and an equivalent dielectric constant $\epsilon_g = 1 + \ii \sigma/\omega\epsilon_0 t_g$. 
For highly-doped MLG, $|\Re \epsilon_g| >> \Im \epsilon_g$ in the local limit ($\omega >> \tau^{-1}$) since $\Im \sigma / \Re \sigma \approx \omega\tau $, providing an intuitive insight to why highly-confined SP modes may be supported by graphene even at THz fequencies.
The variation of the SP dispersion with the number of graphene sheets is shown in Fig.~3(a), with the Fermi-level $\mu$ taken to be $0.9 \eV$. 
By increasing $N$ gradually from 1 to 5, the SP propagation distance increases from about $60 \micron$ to more than $1 \mm$ while its effective wavelength increases from roughly $\lambda/3.2$ to $\lambda/1.6$. 
Figure~3(b) shows that a significant portion of the SP field extends to the neighboring media. 
Of particular interest is the penetration into the gap region ($n_2$) as both the 
width and the angular position of the resonance dip is modified in the presence of the coupling prism~\cite{sambles}. 
The width of the resonance, which is roughly proportional to $\Im k_{sp}$, is broadened by radiative damping of SPs that couple back to the prism as freely-propagating photons. 
For narrower gaps $d$ or where the SP fields are more intense, stronger coupling with the prism also leads to  
a more profound angular shift of the ATR minimum from the angle $\sin^{-1} (n_{sp}/n_p)$. 
It is observed from Fig.~3(b) that for gap distances $d \lesssim 10 \micron$, significant coupling may occur between the SP and the prism. Calculations of the transverse decay length of the evanescent field at the prism/gap interface show that the gap distance should be limited to $d \lesssim 5 \micron$ so that it may couple to the surface mode. 
Therefore, it may be expected that the radiative damping of the SP can affect the ATR minimum non-negligibly.  

\begin{figure}[bp]
\centering
\epsfig{file=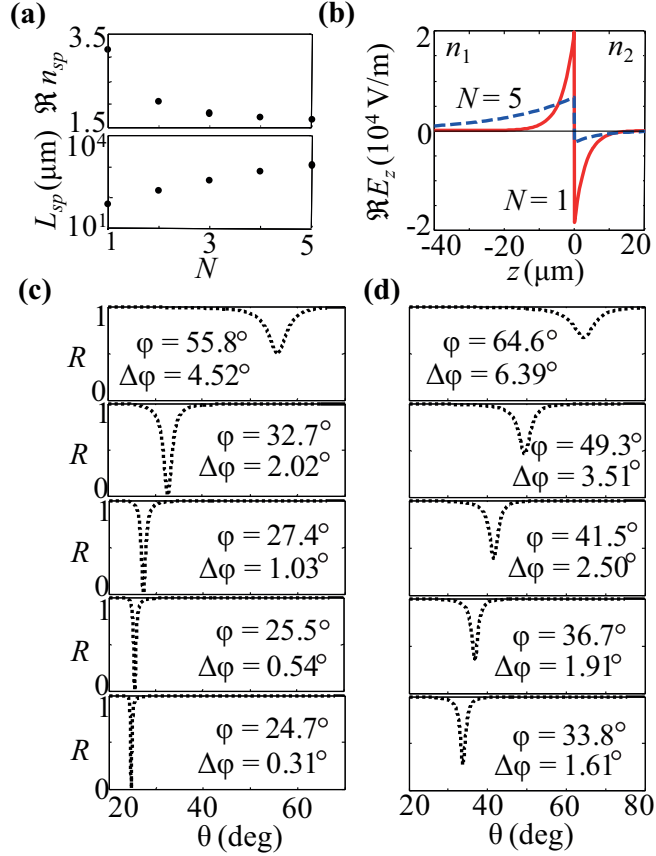, width=3.375 in}
\caption{SP dispersion and simulated angular reflection spectra with MLG and FLG for $f = 5 \THz \,(\hbar\omega = 20.7 \meV)$. (a) SPP dispersion for $N = 1$ to 5. (b) Comparison of the 
normal electric field component $E_z$ for $N = 1$ and $N = 5$. (c) Spectra for $N = 1$ (top) to $N = 5$ (bottom). In (a) to (c), the Fermi-Level $\mu$ is taken to be $0.9 \eV$.    
(d) Spectra for $\mu = \{0.8, 1.0, 1.2, 1.4, 1.6\} \eV$ (top to bottom) for MLG ($N = 1$). For all cases in (c) and (d), the gap distance $d$ was varied between $3 - 6 \micron$.    }
\label{fig:f3}
\end{figure}

The systematic shift in the ATR minimum from $\varphi = 55.8^\circ$ for $N = 1$ to $\varphi = 24.8^\circ$ for $N =5$ is shown in Fig.~3(c), taking the refractive index $n_p \sim 4.0$ for a 
Germanium~\cite{loewenstein} prism. The resonance dip is clearly resolved in each case, with a difference of $0.8^\circ$ between $N = 4$ and $N = 5$. 
For the MLG ($N = 1$), the reflection minimum reaches $50\%$, and the width of the resonance $\Delta\varphi = 4.52^\circ$. For $N \geq 2$, the strength of the SP field that extends into the gap decreases, and sharp narrow dips are obtained with the reduced radiation damping of the SP mode. 
The shift of the ATR minimum from the angle $\sin^{-1} (n_{sp}/n_p)$ is $3.45^\circ$ for the case $N = 1$, and gradually decreases for increasing $N$ to $0.13^\circ$ for $N = 5$. 
Similar trends in the strength and width of the resonance dips are observed in the reflection spectra of Fig.~3(d) for MLG with its Fermi-level increased from $\mu = 0.8 \eV$ to $1.6 \eV \, (N_c \approx 2 \times 10^{14} \cm^{-2})$.   
From the simulated reflection spectra, it is observed that the narrowest and most pronounced resonance dips occur 
for the least lossy SP modes, i.e. for increasing number of layers $N$ or Fermi-level $\mu$. 

It is reasonable to pose the question on the range of frequencies and doping levels for which SPs may be excited on graphene with the Otto configuration. 
Let us revert to the approximate dispersion relation~\eqref{spdispapp} to provide an estimate. Imposing the constraint $\Re n_{sp} < n_p$ leads to the condition 
\eq\label{condit}
\lambda > \delta (n_1^2 + n_2^2)/n_p \,,
\eeq
where 
$\delta = 2 \epsilon_0 (\pi\hbar c)^2/(e^2 \mu)$.
Let us take that the refractive index of the substrate is greater than unity, i.e., $n_1 > 1$. The limit of the inequality expressed in condition~\eqref{condit} is plotted for different values of $n_p$ as a function of the Fermi-level in Fig.~4, taking the 
gap region to be free-space $(n_2 = 1)$. The domain above each curve is where the inequality is satisfied. 
For current achievable levels of doping for graphene~\cite{efetov,fwangnat471},  Fig.~4 shows that the highest frequency for which SPs on MLG may be excited through ATR with high-index coupling prisms is $\sim 10 \THz\, (\lambda \sim 30 \micron)$. For excitation at longer wavelengths (provided $\lambda << 2 \pi c \tau$), heavy doping to increase the Fermi-level becomes essential. 
The condition~\eqref{condit} in terms of photon energy (in units of $\eV$) is $\hbar\omega < e\mu n_p/ \left[\epsilon_0 \pi\hbar c(n_1^2 + n_2^2) \right]$.   
As the inequality holds for MLG, it suggests that excitation frequencies $f \gtrsim 10 \THz \,(\hbar\omega \gtrsim 41.3 \meV)$ may also be considered in the case of FLG.

\begin{figure}[bp]
\centering
\epsfig{file=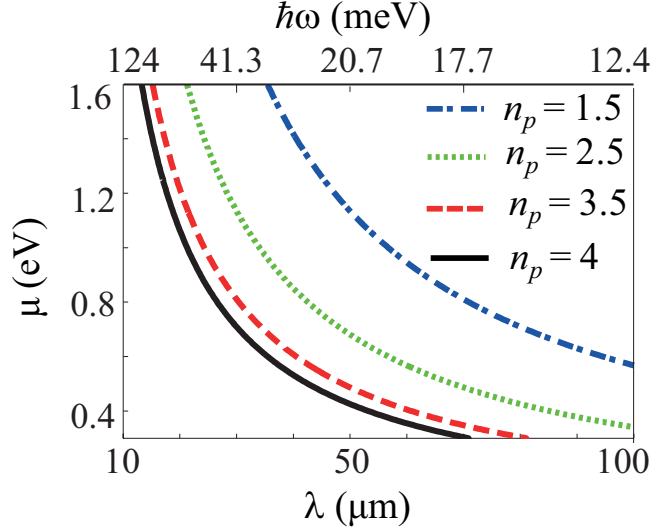, width=3.375 in}
\caption{Estimating the spectral range for which SPs may be excited on MLG with the inequality $\lambda > \delta (n_1^2 + n_2^2)/n_p$ (Eq.~\eqref{condit}), taking the gap region to be free-space ($n_2 = 1$) and $n_1 > 1$. The inequality is satisfied in the domain above each curve. The photon energy (in units of $\meV$) corresponding to the wavelength is shown on the top horizontal axis.}
\label{fig:f4}
\end{figure}

In conclusion, we have demonstrated numerically the excitation of SPs at THz frequencies with highly-doped MLG and FLG via the Otto geometry. In contrast to the noble metals, the SP modes supported on graphene sheets remain highly-confined even in the THz regime. This strong confinement may be explained by considering an equivalent dielectric constant for graphene, from which 
it is apparent that $|\Re \epsilon_g| >> \Im \epsilon_g$.  With rapid advances in the fabrication of MLG and FLG, it is foreseeable that the above simulated results may be realized in a future experiment.
For FLG, in case each of the graphene sheet is separated by a transparent buffer layer ($\sim 20 \nm$ thick) as in Ref.~[17], our calculations 
show that the effect on the position, width, and magnitude on the resonance dips is only marginal. 
As an alternative, the Kretschmann configuration may be employed if the substrate containing the graphene sample may be directly deposited onto the coupling prism. 
If high-index material at THz~\cite{choinat} becomes available, higher-index prisms can be employed to excite SPs on MLG for frequencies beyond $10 \THz$. High-index prisms are also useful to shift the resonance to smaller angles where the transverse decay of the evanescent fields are longer, allowing for larger gap distances $d$. 
Apart from surface plasmon resonance sensors, the proposed configuration may be employed for determining the 
thickness of graphene samples containing MLG and FLG~\cite{ferrarinl,knoll}. Finally, 
it is worth noting that in contrast to the $H-$symmetric mode supported in thin metal films, the field of the $H-$antisymmetric SP mode supported on graphene decays more quickly in the medium of lower refractive index (see Fig.~3(b)). 
As such, the SP mode may be considered for
improving the confinement factor in THz waveguides, for instance in semi-insulating surface plasmon waveguides of the quantum cascade laser~\cite{qhuapl82}. 


\begin{thebibliography}{99}

\bibitem{raether}H. Raether, {\em Surface Plasmons on Smooth and Rough Surfaces and on Gratings}, Springer, Berlin, 1988.

\bibitem{spoof}J. B. Pendry, L. Mart\'{i}n-Moreno L, and F. J. Garcia-Vidal, 
Science {\bf 305}, 847 (2004).


\bibitem{kabashin}A. V. Kabashin, E. Pevans, S. Pastkovsky, W. Hendren, G. A. Wurtz, R. Atkinson, R. Pollard, V. A. Podolskiy, and A. V. Zayats, 
Nat. Materials {\bf 8}, 867 (2009).


\bibitem{ganol}C. H. Gan, and P. Lalanne, 
Opt. Lett. {\bf 35}, 610--612 (2010).



\bibitem{engheta}A. Vakil and N. Engheta, 
Science {\bf 332}, 1291 (2011).

\bibitem{koppens}F. Koppens, D. E. Chang, and F. J. Garc\'{i}a de Abajo, 
Nano Lett. {\bf 11}, 3370 (2011).

\bibitem{ganprb85}C. H. Gan, H. S. Chu, and E. P. Li, 
Phys. Rev. B \textbf{85}, 125431 (2012).

\bibitem{bwangapl100}B. Wang, X. Zhang, X. Yuan, and J. Teng, 
Appl. Phys. Lett. {\bf 100}, 131111 (2012).


\bibitem{novo666}K. S. Novoselov, A. K. Geim, S. V. Morozov, D. Jiang, Y. Zhang, S. V. Dubonos, I. V. Grigorieva, and A. A. Firsov, 
Science {\bf 306}, 666 (2004).

\bibitem{thongapl100}S. Thongrattanasiri, I. Silveiro, and F. J. Garc\'{i}a de Abajo, 
Appl. Phys. Lett. {\bf 100}, 201105 (2012).



\bibitem{junatnano}L. Ju, B. Geng, J. Horng, C. Girit, M. Martin, Z. Hao, H. A. Bechtel, X. Liang, A. Zettl, Y. R. Shen, and F. Wang, 
Nat. Nanotech. {\bf 6}, 1 (2011).


\bibitem{basovnano}Z. Fei, G. Andreev, W. Bao, L. M. Zhang, A. S. McLeod, C.Wang,
M. K. Stewart, Z. Zhao, G.Dominguez, M. Thiemens, M.M. Fogler,
M. J. Tauber, A. H. Castro-Neto, C. N. Lau, F. Keilmann, and D. N. Basov, 
Nano Lett. {\bf 11}, 4701 (2011).


\bibitem{jchenarxiv}J. Chen, M. Badioli, P. Alonso-Gonz\'{a}lez, S. Thongrattanasiri, F. Huth, J. Osmond, M. Spasenovic, A. Centeno, A. Pesquera, P. Godignon, A. Zurutuza, N. Camara, 
J. G. de Abajo, R. Hillenbrand, and F. Koppens, 
Nature {\bf 487}, 77 (2012).


\bibitem{sambles}J. R. Sambles, G. W. Bradbery, and F. Yang, 
Contemporary Phys. {\bf 32}, 173 (1991).

\bibitem{fratiniprb77}S. Fratini, and F. Guinea, 
Phys. Rev. B {\bf 77}, 195415 (2008).


\bibitem{ferrarinl}C. Casiraghi, A. Hartschuh, E. Lidorikis, H. Qian, H. Harutyunyan, T. Gokus, K. S. Novoselov, and A. C. Ferrari, 
Nano Lett. {\bf 7}, 2711 (2007).

\bibitem{xianatnano}H. Yan, X. Li, B. Chandra, G. Tulevski, Y. Wu, M. Freitag, W. Zhu, P. Avouris, F. Xia, 
Nat. Nanotech. {\bf 7}, 330 (2012).


\bibitem{ranaapl93}J. M. Dawlaty, S. Shivaraman, J. Strait, P. George, M. Chandrashekhar, F. Rana, M. G. Spencer, D. Veksler, and Y. Chen,
Appl. Phys. Lett. {\bf 93}, 131905 (2008).

\bibitem{hansonjap103}G. Hanson, 
J. Appl. Phys. {\bf 103}, 064302 (2008).

\bibitem{jablanprb80}M. Jablan, H. Buljan, and M. Solja\v{c}ic, 
Phys. Rev. B {\bf 80}, 245435 (2009).



\bibitem{efetov}D. K. Efetov and P. Kim, 
Phys. Rev. Lett. {\bf 105}, 256805 (2010).

\bibitem{fwangnat471}C. Chen, C. Park, B. W. Boudouris, J. Horng, B. Gene, C. Girit, A. Zettl, M. F. Crommie, 
R. A. Segalman, S. G. Louie, and F. Wang, 
Nature {\bf 471}, 617 (2011).


\bibitem{yariv}A. Yariv, and P. Yeh, {\em Optical Waves in Crystals}, John Wiley \& Sons, New Jersey, 2003.

\bibitem{loewenstein}E. V. Loewenstein, D. R. Smith, and R. L. Morgan, 
Appl. Opt. \textbf{12} (1973), 398.


\bibitem{choinat}M. Choi, S. H. Lee, Y. Kim, S. B. Kang, J. Shin, M. H. Kwak, K.-Y. Kang, Y.-H. Lee, N. Park, and B. Min, 
Nature {\bf 470}, 369 (2011).

\bibitem{knoll}W. Hickel, G. Duda, M. Jurich, T. Kr\"{o}hl, K. Rochford, G. I. Stegeman, J. D. Swalen, G. Wegner, and W. Knoll, 
Langmuir {\bf 6}, 1403 (1990).


\bibitem{qhuapl82}B. S. Williams, H. Callebaut, S. Kumar, Q. Hu, and J. L. Reno, 
Appl. Phys. Lett. {\bf 82}, 1015 (2003).



\end{thebibliography}
\end{document}